\documentclass[journal, comsoc, final]{IEEEtran}

\usepackage{graphicx}
\usepackage{blindtext}
\usepackage{subfig}

\usepackage{pifont}
\usepackage{cite}
\usepackage{amsmath}
\usepackage{float}
\usepackage{multirow}
\usepackage[hyphens]{url}

\usepackage{array}
\usepackage{tabularx}
\newcolumntype{P}[1]{>{\centering\arraybackslash}p{#1}}
\newcolumntype{M}[1]{>{\centering\arraybackslash}m{#1}}
\newcolumntype{L}[1]{>{\centering\arraybackslash}l{#1}}
\newcolumntype{Z}{>{\centering\let\newline\\\arraybackslash\hspace{0pt}}X}

\newcolumntype{Y}[1]{>{\raggedright\let\newline\\\arraybackslash\hspace{0pt}}X{#1}}

\begin{document}

\title{SDN Controllers: Benchmarking \& Performance Evaluation}

\author{
	
	Liehuang~Zhu,~\IEEEmembership{Member,~IEEE},
	Md Monjurul Karim,
	Kashif~Sharif,~\IEEEmembership{Member,~IEEE},
	Fan~Li,~\IEEEmembership{Member,~IEEE},
	Xiaojiang~Du,~\IEEEmembership{Senior~Member,~IEEE},
	and Mohsen~Guizani,~\IEEEmembership{Fellow,~IEEE}
	
	\thanks{
		L. Zhu, M. Karim, K. Sharif, and F. Li are with Beijing Engineering Research Center for Massive Language Information Processing \& Cloud Computing Application, and School of Computer Science and Technology, Beijing Institute of Technology, Beijing, China. Email: \{liehuangz,mkarim,kashif,fli\}@bit.edu.cn		
	}
	\thanks{Xiaojiang Du is with Department of Computer and Information Sciences, Temple University, Philadelphia, USA. Email: dux@temple.edu}
	\thanks{Mohsen Guizani is with Department of Electrical and Computer Engineering, University of Idaho, Moscow, ID, USA. Emil: mguizani@uidaho.edu}
	\thanks{K. Sharif is the corresponding author.}
}

\markboth{A version is under review at IEEE JSAC}{}
\maketitle

\begin{abstract}
Software Defined Networks offer flexible and intelligent network operations by splitting a traditional network into a centralized control plane and a programmable data plane. The intelligent control plane is responsible for providing flow paths to switches and optimizes network performance. The controller in the control plane is the fundamental element used for all operations of data plane management. Hence, the performance and capabilities of the controller itself are extremely important. Furthermore, the tools used to benchmark their performance must be accurate and effective in measuring different evaluation parameters. There are dozens of controller proposals available in existing literature. However, there is no quantitative comparative analysis for them. In this article, we present a comprehensive qualitative comparison of different SDN controllers, along with a quantitative analysis of their performance in different network scenarios. More specifically, we categorize and classify 34 controllers based on their capabilities, and present a qualitative comparison of their properties. We also discuss in-depth capabilities of benchmarking tools used for SDN controllers, along with best practices for quantitative controller evaluation. This work uses three benchmarking tools to compare nine controllers against multiple criteria. Finally, we discuss detailed research findings on the performance, benchmarking criteria, and evaluation testbeds for SDN controllers. 
\end{abstract}


\section{Introduction}
\label{sec:intro}

\begin{figure}[!b]
	\centering
	\includegraphics[width=0.8\linewidth]{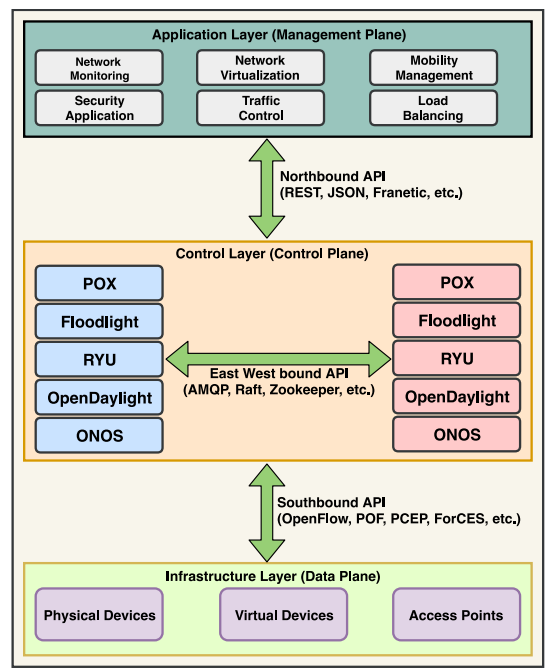}
	\caption{Elements in a layered structure of SDN.}
	\label{fig:architecture_sdn}
\end{figure}

\IEEEPARstart{S}{oftware} Defined Networks (SDN) have seen tremendous growth and deployment in different types of networks in recent times. 
They are being actively used in datacenter networks \cite{jain2013b4,hong2018b4}, wireless \& Internet of Things (IoT) networks \cite{bera2017software,haque2016wireless}, wide area \& cellular networks. \cite{nguyen2017sdn}, as well as security and privacy of domains\cite{ZhuTang}.
Compared to traditional networks it decouples the control logic from network layer devices, and centralizes it for efficient traffic forwarding and flow management across the domain. This multi-layered architecture, as shown in Figure~\ref{fig:architecture_sdn}, has data forwarding devices at the bottom in data plane, which are programmed by controllers in the control plane. The high level application or management plane interacts with control layer to program the whole network and enforce different policies. The interaction among these layers is done through interfaces which work as communication/programming protocols.

Traditional networks suffer from a number of limitations, mainly due to diverse service requirements and the scale of the network. Some of these are related to traffic engineering, flow management, policy enforcement, security, and virtualization~\cite{A,B,C,D,E}. 
SDN presents a simplified, centralized, and efficient solution to these, by decoupling the data plane forwarding and control plane intelligence. Hence, the network switched become simple forwarding devices, which route data traffic based on instruction from a softwarized controller.
This centralized entity provides a programmatic control of whole network and enables real-time control of underlying devices. By using SDN, network management becomes straightforward and helps in removing rigidity from the network.

Some of the well known controllers are NOX \cite{Gude2008}, POX \cite{GitHubPOX}, Floodlight \cite{BigSwitchNetworks}, OpenDaylight (ODL) \cite{web-opendaylight}, Open Network Operating System (ONOS) \cite{Berde2014} and RYU \cite{GitHubRyu}. However, a number of other controllers and flavors are available in the literature. From a practical implementation perspective, it is very difficult to determine which controller will perform best in any given type of network. Hence, the qualitative and quantitative comparative analysis of these controllers is very important. To the best of our knowledge, there is no such work which compares the controllers for their properties and evaluates their performance. Although a number of surveys have been done for SDN in general, there are none which provide a comprehensive controller evaluation. Works in \cite{tootoonchian2012controller,shah2013architectural,shalimov2013advanced,jarschel2012flexible,jarschel2014ofcprobe,Khattak2014,khondoker2014feature,zhao2015performance,isaia2016performance,Mallon2016,salman2016sdn,suh2017toward,bholebawa2018performance,nguyen2018benchmarking} present some quantitative comparison, however, most of them either feature a specific application or simple environment to execute multiple experiments. 
In this work, we have adopted a different method by using a number of different benchmarking tool specifically developed for controller evaluations. The contributions of this work are multi-fold:
\begin{itemize}
	\item We present the generic architecture of SDN controller and the evolution of modern SDN controllers.
	\item We present a qualitative comparative analysis of 34 different controllers for their properties and capabilities. We also discuss the different use cases for these controllers and the enhancements done to improve their performance by other works.
	\item We present a comprehensive study of benchmarking techniques and tools for SDN controllers. This includes the existing works \& approaches used for evaluation, capabilities of benchmarking tools, and most importantly the details of metrics which should be used for quantitative evaluations.
	\item We conduct quantitative analysis of 9 different controllers using 3 different benchmarking tools for a variety of metrics. The results presented show the actual performance of controllers.
	\item We present comprehensive discussion on research findings not only for controller behavior but also for the metrics and tools used.
\end{itemize}

The rest of the paper is organized as follows: Section II gives and overview of SDN controllers, followed by comparison and classification of controllers in Section III. Benchmarking metrics and existing efforts are detailed in Section IV. Benchmarking tools and their properties are evaluated in Section V. Experimental results and research findings are detailed in section VI and VII respectively. Section VIII concludes the paper.

\section{SDN Controllers}
A controller is the core component of any SDN infrastructure, as it has the global view of entire network including data plane SDN devices. It connects these resources with management applications, and performs flow actions dictated by application policy among the devices.
In this section, we present the generic architecture of the controllers, and the evolution towards modern controllers. We also present the classification, comparison, and use case enhancements for 34 different controllers.

\begin{figure}[!t]
	\centering
	\includegraphics[width=0.85\linewidth]{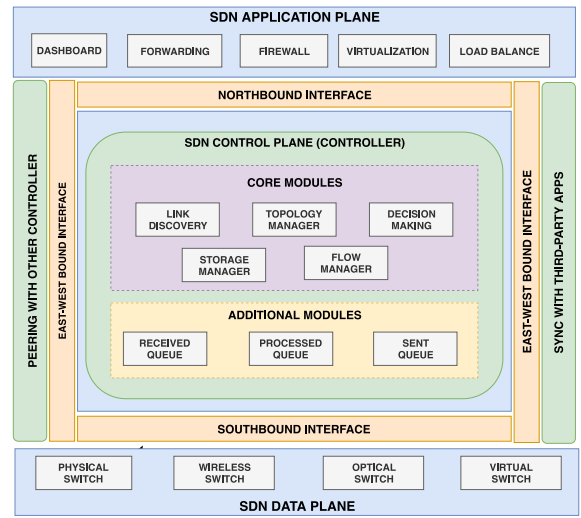}
	\caption{General Overview of SDN Controller}
	\label{fig:controller_architecture}
\end{figure}

\subsection{Architecture of SDN Controllers}
The controller in a software defined network, also referred as Network Operating Systems (NOS), is the core and critical component responsible for making decisions on managing traffic in underlying network.
The proposals put forth for different controllers in literature do not modify the basic controller architecture, rather they differ in terms of modules and capabilities. Hence, we find that presenting individual architectures to be less useful for the reader. Here, we present the general architecture as shown in Figure~\ref{fig:controller_architecture}, and discuss its different modules.

\textbf{Controller Core:} The core functions of the controller are mainly related to topology and traffic flow. The link discovery module regularly transmits inquiries on external ports utilizing packet\_out messages. These inquiry messages return in the from of packet\_in messages, which allows the controller to build the topology of network. The topology itself is maintained by the topology manager. This provides the decision making module to find optimal paths between nodes of the network. The paths are built such that the different QoS policies or security policies can be enforced during path installation. In addition, the controller may also have dedicated statistics collector/manager and queue manager for collecting performance information and management of different incoming and outgoing packet queues, respectively. Flow manager is one of the major modules which directly interacts with data plane's flow entries and flow tables.  It utilizes southbound interface for this purpose.

\textbf{Interfaces:} The core controller is surrounded by different interfaces for interaction with other layers and devices.
Southbound Interface (SBI) defines a set of processing rules that enable packet forwarding between forwarding devices and controllers. SBI helps the controller to provision physical and virtual network devices intelligently. OpenFlow (OF) \cite{McKeown2008} is the most commonly used SBI and is a de-facto standard for industry. The fundamental responsibility  of OF is to define flows and classify network traffic based on a predefined rule set.
On the opposite end, the controller uses Northbound Interface (NBI) to allow developers to integrate their applications with controller and data plane devices. Controllers support a number of northbound APIs, but most of them are based on REST API.
For inter controller communication, West Bound Interface (WBI) is used. There is no standard communication interface for this purpose, hence different controllers use different mechanisms. Moreover, heterogeneous controllers do not usually communicate with each other. East Bound API (EBI) extends the capability of controller to interact with legacy routers. BGP \cite{rekhter2006} is the most commonly used protocol for this purpose.

\subsection{Evolution of SDN Controllers}
Modern SDN controllers and SDN design is not the first attempt at centralizing the network control. From mid-2000s, several attempts have been made to separate the control logic from the data plane.

SoftRouter~\cite{lakshman2004softrouter} and ForCES~\cite{dantu2004forwarding} were introduced in a single network device to separate control elements~(CEs) from forwarding elements~(FEs). However, they were limited to packet modification functionalities, as most of the routers (at the time) were limited in computing intelligence or network awareness to perform required operations. Routing Control Platforms~(RCP)~\cite{feamster2004case} was proposed as an intra-AS (Autonomous System) platform to implement an expandable control platform for BGP. However, the solution is for heterogeneous networks and prone to single point of failure. Path Computation Engine~(PCE)~\cite{farrel2006} was presented to enable clients to execute path computations in routers but lacks dedicated centralized path computation engine and fails to provide cooperation among different entities. Although Intelligent Route Service Control Point (IRSCP)~\cite{van2006dynamic} introduces path allocation module in an external router and provides dynamic connectivity feature to enhance traffic flows throughout a network, it was limited to single ISP service. On the other hand, 4D project~\cite{greenberg2005clean} was intended as a clean-state solution to introduce a control plane for topology discovery and to provide traffic forwarding logic and rule sets. However, there is no practical implementation of this approach. The SANE project~\cite{casado2006sane} was developed by National Science Foundation (NSF) to enable traffic forwarding and access control policies using logically centralized server within enterprise networks. Ethane~\cite{casado2007ethane} is the successor of the SANE project that brings a more improved and practical control management module, aware of global network and performs routing operations based on pre-defined flows. Both, SANE and Ethane fail to acknowledge the network components as an overall representation. Besides, they also lack flow-level control over traditional routing approaches. 

The control plane of these earlier proposals is missing a broad range of matching header fields and also lacks a wide range of functionalities. As a result, SDN has become mainstream with the introduction of OpenFlow \cite{McKeown2008} which is a data-plane Application Programming Interface (API), and a robust centralized controller named NOX \cite{Gude2008}.  OpenFlow is different from previous solutions as it is an open protocol to favor software developers to build applications on different switches that support flow tables with an extensible range of header fields. SDN brings flexibility and agility by allowing virtualization of the servers, rapid response to network changes, deployment of policies, and centralized control over complete network.

\section{Classification and Comparison of SDN Controllers}
In order to compare different SDN controllers, we have performed an extensive search of proposals not only in academic literature, but also in commercial domain. Here, we first present the possible classification criteria of controllers, followed by the comparative analysis, and then different use case specific enhancements.

\begin{table*}[!t]
	\centering
	\caption{SDN Controller's Feature Comparison Table}
	\label{table_controller_compare}
	\tiny
	\setlength\tabcolsep{2pt}
	\begin{tabularx}
		{\linewidth}{|>{\hsize=\hsize}Y|>{\hsize=\hsize}Z|>{\hsize=\hsize}Z|>{\hsize=\hsize}Z|>{\hsize=\hsize}Z|>{\hsize1\hsize}Z|>{\hsize=\hsize}Z|>{\hsize=\hsize}Z|>{\hsize=\hsize}Z|>{\hsize=\hsize}Z|>{\hsize=\hsize}Z|>{\hsize=\hsize}Z|>{\hsize=\hsize}Z|}\hline
		
		\textbf{Name}           & \textbf{Programming Language} & \textbf{Architecture}             & \textbf{Northbound API}                & \textbf{Southbound API}                    & \textbf{EastWestbound API}              & \textbf{Supported Platform}    & \textbf{Interface}                                            & \textbf{License}     & \textbf{Multithreading} & \textbf{Modularity} & \textbf{Consistency} & \textbf{Documentation}\\ 
		\hline
		
		Beacon\cite{Erickson2013}         & Java                 & Centralized              & ad-hoc                        & OpenFlow 1.0                      & -                              & Linux, MacOS, Windows & CLI, Web UI                                          & GPL 2.0     & Yes            & Fair       & No          & Fair           \\\hline

		Beehive\cite{Yeganeh2014}        & Go                   & Distributed Hierarchical & REST                          & OpenFlow 1.0, 1.2                 & -                              & Linux                 & CLI                                                  & Apache 2.0  & Yes            & Good       & Yes         & Limited        \\ 		\hline
		
		DCFabric \cite{DCFabricGitHub}       & C, Javascript        & Centralized              & REST                          & OpenFlow 1.3                      & -                              & Linux                 & CLI, Web UI                                          & LGPL 3.0    & Yes            & Good       & Yes         & Fair           \\ 		\hline
		
		Disco \cite{Phemius2014}          & Java                 & Distributed Flat         & REST                          & OpenFlow 1.0                      & AMQP                           & -                     & -                                                    & Proprietary & -              & Good       & No           & Limited        \\ 		\hline
		
		Faucet \cite{bailey2016faucet}         & Python               & Centralized              & -                             & OpenFlow 1.3                      & -                              & Linux                 & CLI, Web UI                                          & Apache 2.0  & Yes            & -          & Yes         & Good           \\ 		\hline
		
		Floodlight \cite{BigSwitchNetworks}     & Java                 & Centralized              & REST, Java RPC, Quantum       & OpenFlow 1.0, 1.3                 & -                              & Linux, MacOS, Windows & CLI, Web UI                                          & Apache 2.0  & Yes            & Fair       & Yes         & Good           \\ 		\hline
		
		FlowVisor \cite{Sherwood2009b}      & C                    & Centralized              & JSON RPC                      & OpenFlow 1.0, 1.3                 & -                              & Linux                 & CLI                                                  & Proprietary & -              & -          & No          & Fair           \\ 		\hline
		
		HyperFlow \cite{Tootoonchian2010}      & C++                  & Distributed Flat         & -                             & OpenFlow 1.0                      & Publish and subscribe messages & -                     & -                                                    & Proprietary & Yes            & Fair       & No          & Limited        \\ 		\hline
		
		Kandoo \cite{HassasYeganeh2012}         & C, C++, Python       & Distributed Hierarchical & Java RPC                      & OpenFlow 1.0-1.2                  & Messaging Channel              & Linux                 & CLI                                                  & Proprietary & Yes            & High       & No          & Limited        \\ 		\hline
		
		Loom \cite{Kazarez}           & Erlang               & Distributed Flat         & JSON                          & OpenFlow 1.3-1.4                  & -                              & Linux                 & CLI                                                  & Apache 2.0  & Yes            & Good       & No           & Good           \\ 		\hline
		
		Maestro \cite{Cai2011}        & Java                 & Centralized              & ad-hoc                        & OpenFlow 1.0                      & -                              & Linux, MacOS, Windows & Web UI                                               & LGPL 2.1    & Yes            & Fair       & No          & Limited        \\ 		\hline
		
		McNettle \cite{Voellmy2012}       & Haskell              & Centralized              & -                             & OpenFlow 1.0                      & -                              & Linux                 & CLI                                                  & Proprietary & Yes            & Good       & No          & Limited        \\ 		\hline
		
		Meridian \cite{Banikazemi2013}       & Java                 & Centralized              & REST                          & OpenFlow 1.0, 1.3                 & -                              & Cloud-based           & Web UI                                               & -           & Yes            & Good       & No           & Limited        \\ 		\hline
		
		Microflow \cite{MicroFlow-Web}      & C                    & Centralized              & Socket                        & OpenFlow 1.0-1.5                  & -                              & Linux                 & CLI, Web UI                                          & Apache 2.0  & Yes            & -          & No          & Limited        \\ 		\hline
		
		NodeFlow \cite{nodeflowweb}       & JavaScript           & Centralized              & JSON                          & OpenFlow 1.0                      & -                              & Node.js               & CLI                                                  & Cisco       & -              & -          & No          & Limited        \\ 		\hline
		
		NOX \cite{Gude2008}            & C++                  & Centralized              & ad-hoc                        & OpenFlow 1.0                      & -                              & Linux                 & CLI, Web UI                                          & GPL 3.0     & Yes (Nox-MT )  & Low        & No          & Limited        \\ 		\hline
		
		Onix \cite{Koponen2010}           & C++                  & Distributed Flat         & Onix API                      & OpenFlow 1.0, OVSDB               & Zookeeper                      & -                     & -                                                    & Proprietary & Yes            & Good       & No           & Limited        \\ 		\hline
		
		ONOS \cite{Berde2014}           & Java                 & Distributed Flat         & REST, Neutron                 & OpenFlow 1.0, 1.3                 & Raft                           & Linux, MacOS, Windows & CLI, Web UI                                          & Apache 2.0  & Yes            & High       & Yes         & Good           \\ 		\hline
		
		OpenContrail \cite{opencontrailwebpage}   & C, C++, Python       & Centralized              & REST                          & BGP, XMPP                         & -                              & Linux                 & CLI, Web UI                                          & Apache 2.0  & Yes            & High       & Yes         & Good           \\ 		\hline
		
		OpenDaylight \cite{web-opendaylight}   & Java                 & Distributed Flat         & REST, RESTCONF, XMPP, NETCONF & OpenFlow 1.0, 1.3                 & Akka, Raft                     & Linux, MacOS, Windows & CLI, Web UI                                          & EPL 1.0     & Yes            & High       & Yes         & Good           \\ 		\hline
		
		OpenIRIS \cite{Lee2014}       & Java                 & Distributed Flat         & REST                          & OpenFlow 1.0-1.3                  & Custom Protocol                & Linux                 & CLI, Web UI                                          & Apache 2.0  & Yes            & Fair       & No           & Limited        \\ 		\hline
		
		OpenMul \cite{openmul-web}        & C                    & Centralized              & REST                          & OpenFlow 1.0, 1.3, OVSDB, Netconf & -                              & Linux                 & CLI                                                  & GPL 2.0     & Yes            & High       & No           & Good           \\ 		\hline
		
		PANE \cite{Ferguson2013}           & Haskell              & Distributed Flat         & PANE API                      & OpenFlow 1.0                      & Zookeeper                      & Linux, MacOS          & CLI                                                  & BSD 3.0     & -              & Fair       & No          & Fair           \\ 		\hline
		
		POF Controller \cite{li2017protocol} & Java                 & Centralized              & -                             & OpenFlow 1.0, POF-FIS             & -                              & Linux                 & \begin{tabular}[c]{@{}c@{}}CLI, \\ GUI \end{tabular} & Apache 2.0  & -              & -          & No          & Limited        \\ 		\hline
		
		POX \cite{GitHubPOX}            & Python               & Centralized              & ad-hoc                        & OpenFlow 1.0                      & -                              & Linux, MacOS, Windows & CLI, GUI                                             & Apache 2.0  & No             & Low        & No          & Limited        \\ 		\hline
		
		Ravel \cite{Wang2016}          & Python               & Centralized              & ad-hoc                        & OpenFlow 1.0                      & -                              & Linux                 & CLI                                                  & Apache2.0   & -              & -          & Yes         & Fair           \\ 		\hline
		
		Rosemary \cite{Shin2014}       & C                    & Centralized              & ad-hoc                        & OpenFlow 1.0, 1.3, XMPP           & -                              & Linux                 & CLI                                                  & Proprietary & Yes            & Good       & No          & Limited        \\ 		\hline
		
		RunOS \cite{RunOS-Web}          & C++                  & Distributed Flat         & REST                          & OpenFlow 1.3                      & Maple                          & Linux                 & CLI, Web UI                                          & Apache2.0   & Yes            & High       & Yes         & Fair           \\ 		\hline
		
		Ryu \cite{GitHubRyu}            & Python               & Centralized              & REST                          & OpenFlow 1.0-1.5                  & -                              & Linux, MacOS          & CLI                                                  & Apache 2.0  & Yes            & Fair       & Yes         & Good           \\ 		\hline
		
		SMaRtLight \cite{Botelho2014a}     & Java                 & Distributed Flat         & REST                          & OpenFlow 1.3                      & BFT-SMaRt                      & Linux                 & CLI                                                  & Proprietary & -              & -          & No           & Limited        \\ 		\hline
		
		TinySDN \cite{TrevizanDeOliveira2015}        & C                    & Centralized              & -                             & OpenFlow 1.0                      & -                              & Linux                 & CLI                                                  & BSD 3.0     & No             & -          & No          & Limited        \\ 		\hline
		
		Trema \cite{takamiya2012}          & C, Ruby              & Centralized              & ad-hoc                        & OpenFlow 1.0                      & -                              & Linux                 & CLI                                                  & GPL 2.0     & -              & Good       & No           & Fair           \\ 		\hline
		
		Yanc \cite{Michel}           & C, C++               & Distributed Flat         & REST                          & OpenFlow 1.0-1.3                  & yanc File System               & Linux                 & CLI                                                  & Proprietary & -              & -          & No          & Limited        \\ 		\hline
		
		ZeroSDN \cite{Kohler}        & C++                  & Distributed Flat         & REST                          & OpenFlow 1.0, 1.3                 & ZeroMQ                         & Linux                 & CLI, Web UI                                          & Apache 2.0  & -              & High       & Yes         & Fair           \\		\hline
	\end{tabularx}
\end{table*}

\subsection{Classification \& Selection Criteria}
The working of controllers is more or less same across all the proposals listed in Table~\ref{table_controller_compare}. After analysis of 34 controllers we conclude that the working, role, and responsibilities of majority of these do not present any classification basis. Perhaps the only classification criteria that can be used is the \emph{deployment architecture}. The initial aim of SDN was to centralize the control plane, hence most of the controllers utilized a single controller, however, this created single point of failure and scalability challenges. The distributed architecture allows usage of multiple controllers inside a domain, working in a flat or hierarchical formation.

In this work, we have not limited the selection of controllers to any specific criteria. Rather we have collected all possible controllers from literature and other documented projects. To the best of our knowledge, there is no other work that collects and compares such a large number of controllers. 

\subsection{Qualitative Comparison}
Table~\ref{table_controller_compare} presents a comprehensive view of different properties of the controllers. In the interest of space and the fact that not all proposals provide extensive details about their inner-workings, we do not discuss each controller individually. Rather we present the properties and design choices of controllers. 

\textbf{Programming Language:}
Controllers have been written using different programming languages, such as C, C++, Java, Java Script, Python, Ruby, Haskell, Go, and Erlang. In some cases, the entire controller is built using a single language. While in many other controllers multiple languages are used in their core and modules, so that they can offer efficient memory allocation, can be executed on multiple platforms, or most importantly achieve higher performance under certain conditions.

\textbf{Architecture:}
The major design decision of a controller is its architecture, which can be centralized or distributed. Centralized controllers are mostly used in small scale networks, whereas distributed controllers are able to span across multiple domains. They can further be classified into flat, where all controller instances have equal responsibilities, or hierarchical, where a root controller is present.

\textbf{Programmable Interface (API):}
Generally, Northbound API (NBI) allows the controller to facilitate applications like topology monitoring, flow forwarding, network virtualization, load-balancing, and intrusion detection based on the network events which are generated by data plane devices. On the other hand, low-level API like Southbound API (SBI) is responsible for enabling the communication between a controller and SDN enabled switches or routers. Additionally, east-west API (EWBI) is used by multiple controllers from different domains to form peering with each other in a distributed or hierarchical environment. Not all controllers provide all APIs, and only select few have customized them for their own specific use.

\textbf{Platform and Interface:}
These properties describe the implementation of controller to be compatible with specific operating system. Majority of controllers are built on top of Linux distributions. Moreover, in order to configure and view statistical information, some controllers provide graphical or web based interfaces to the administrators.

\textbf{Threading and Modularity:}
A single-threaded controller is more suitable for lightweight SDN deployments. In contrast, multi-threaded controllers are suitable for commercial purposes such as, 5G, SDN-WAN, and optical networks. On the other hand, a controller's modularity allows the integration of different applications and functionalities. High modularity allows a controller to perform faster task execution in a distributed environment. 

\textbf{License, Availability, and Documentation:}
Most of the controllers discussed in this article are licensed as Open-Source. However, a few have a proprietary license which means they are only available through special request or for research purpose. Regular maintenance of these controllers is also a challenging task for the developers which is why a number of them do not receive regular updates. Nevertheless, the source code is available online which allows anyone to make further changes according to the requirements.  While accessing them online, we have found that the majority of them lack proper documentation. On the contrary, the ones which are updated on a regular basis feature detailed and updated documentation for all the available version and also include community-based support. 

\subsection{Use case Specific Enhancements to SDN Controllers}
The adoption of different controllers and SDN in general, has also triggered enhancements and use case specific improvements for different controllers. Here, we have grouped these enhancements into different categories, and summarize how they improve the capabilities of controllers.

\subsubsection{Network Monitoring}
Network monitoring has become one of the most vital use cases of SDN controllers. SDN controller can take advantage of the global view of topology and proactively query the performance. OpenTM~\cite{tootoonchian2010opentm} was proposed by as a module for NOX, one of the earliest open-source OpenFlow controller. This monitoring scheme evaluates Traffic Matrix (TM) of OpenFlow switches with a consistent polling rate. However, this also leads to higher monitoring overhead. Adrichem et al. \cite{van2014opennetmon} presented OpenNetMon, a Python-based module for POX controller to monitor end-to-end per flow QoS metrics like throughput, delay, packet loss, etc. From the statistical analysis results, the approach for monitoring throughput is excellent, although continuous polling of information make cause overhead on the controller. Flow monitoring is limited to edge switches only. On the other hand, Payless \cite{chowdhury2014payless} implemented over Floodlight controller is another query-based monitoring framework that can request the desired QoS metrics using a set of well-defined RESTful APIs. However, some trade-off between accuracy and overhead can lead to slight performance degradation for different polling intervals. SDN Interactive Manager \cite{isolani2015sdn} and OFMon \cite{kim2016ofmon} are two recent implementation of network monitoring modules that have been built over Floodlight and ONOS controller respectively.

\subsubsection{Load Balancing}
SDN controller plays an important role to enable load balancing in distributed systems by optimizing resource allocation, minimizing response time, and maximizing throughput of that system. Without rewriting IP addresses, Handigol et al. \cite{handigol2009plug} implemented a method where NOX controller can be used along with OpenFlow switch reactively to reduce response time for load balancing of multiple web servers. Contrarily, Uppal et al. \cite{uppal2010openflow} used address rewriting techniques for NOX-based load balancer which cuts down cost and brings flexibility. Another NOX-based proactive load balancer was proposed by Wang et al. \cite{wang2011openflow} which uses OpenFlow wild card rules that can achieve faster adaptation with new load balancing weights and to redistribute the existing weight more efficiently. Based on switch migration technique Liang at el. in \cite{liang2014scalable} presented a dynamic load balancing method that has been implemented over cluster OpenDaylight controller \cite{web-opendaylight}. However, this method may fail in large scale networks due to coordinator node's recurring load collection issue. 

\subsubsection{Network Virtualization \& Cloud Orchestration}
With addition of Network virtualization (NV) techniques SDNs have gained a new dimension. This has allowed network slicing and multi-tenant hosting on existing physical network resources. FlowVisor \cite{Sherwood2009b} is the most popular SDN based implementation to utilize virtual networks by leveraging OpenFlow functionality to abstract the underlying hardware. VeRTIGO \cite{corin2012vertigo} is an extension of FlowVisor that provides the controllers to choose the depth of virtual network abstraction  required. This extension increases more flexibility in provisioning SDNs, however at the cost of hypervisor complexity. in order to reduce complexity of network management, Xingtao et al. \cite{xingtao2016network} presented an SDN controller built on docker \cite{dockerwiki} to improve the deployment speed with expanded mobility. In \cite{drutskoy2013scalable} the flexibility of NOX controller has been used as a container-based controller virtualization module to effectively cache and manage mappings between virtual networks and physical switches. HyperFlex \cite{blenk2015hyperflex} proposes a control plane virtualization model which largely aims at achieving scalability, privacy, and extensibility. In this architecture, FlowVisor and Ryu controllers have been combined to provide the core hypervisor functions and to control the hypervisor network respectively.

Cloud orchestration defines the integration of SDN controllers with a cloud based resource manager, such as OpenStack \cite{webpage-openstack} to enable dynamic interworking between data centers, wide area networks, transport network, and other enterprise networks. In \cite{mayoral2017sdn}, OpenDaylight is integrated with OpenStack Havana \cite{web-openstack-havana} to evaluate the effectiveness of SDN in a cloud-based architecture where multiple data centers (DC) are located in different domains. In this architecture, the controller communicates with Havana using its REST NBI to perform critical tasks such as building, removal, and migration of virtual instances which are located in inter-DC and intra-DC environments. 

\subsubsection{Policy Enforcement}
To enhance the security and flexible network management, an SDN controller has the capability to assign different policy decisions by implementing flow-based forwarding rules. Hinrichs et al. \cite{hinrichs2008expressing} implemented NOX as an application to provide access control, external authentication, and to enable policy enforcement along with network isolation. PANE \cite{Ferguson2013} presents an API to allow administrators to install policies for bandwidth allocation, access control,  and path control. Additionally, the API provides the capability to query the state of network or to provide information to SDN controller regarding future traffic characteristics.  PolicyCop \cite{bari2013policycop} based on Floodlight controller, is an autonomic QoS policy enforcement architecture, that presents an interface for specifying QoS requirements in Service Layer Arguments and implementation through the OpenFlow API. Besides, it can monitor different policies so that control plane rules can be modified with changing traffic conditions autonomously. An extra module of ONOS controller has been extended to implement a policy-based secure framework in \cite{varadharajan2017securing}. The authors allowed an end-to-end SDN services across various domains including inter and intra domain, using a wild card based policy language which includes a group of entities and services. Associated action such as acceptance or denial of a request is executed when a policy statement is satisfied. 

\section{Benchmarking Process \& Metrics}
Theoretical comparison based on features and properties do not reflect the actual performance of any controller. Hence, real deployment and benchmarking is necessary for true evaluation.
In this section, we first present and overview on the necessity and importance of evaluating controllers. Following it, we discuss existing efforts for benchmarking along with important lessons learned. Finally, we present a list of performance metrics, which should be used in benchmarking of controllers.

\begin{table*}[!h]
	\centering
	\caption{Comparative analysis of different benchmarking studies.}
	\label{tab:Related_Comparison}
	\tiny
	\setlength\tabcolsep{2pt}
	\begin{tabularx}
		{\linewidth}{|>{\hsize=0.4\hsize}Z|>{\hsize=1.3\hsize}Y|>{\hsize=0.5\hsize}Z|>{\hsize=0.9\hsize}Y|>{\hsize=1.1\hsize}Y|>{\hsize=1\hsize}Y|>{\hsize=1.8\hsize}Y|}\hline
		
		\textbf{Reference} & \textbf{Testbed Specifications} & \textbf{Evaluation Tool Used} & \textbf{Controller(s) Evaluated} & \textbf{Evaluation Metrics} & \textbf{Optimization Objectives} & \textbf{Lessons Learned}\\\hline
		
		\cite{tootoonchian2012controller} & 
		1 $\times$ Quad-core \& 1 $\times$ Octa-Core Server\newline
		2 Gbps Link Speed	&
		CBench	&
		NOX, NOX-MT, Beacon, Maestro	&
		Throughput, Latency	&
		Batching I/O\newline Boost Async I/O	&
		Number of switches impact the controller performance. \\\hline
		
		\cite{shah2013architectural} &
		1 $\times$ Cluster with 2 Separate Xeon Servers\newline
		8 Gbps Link Speed	&
		CBench	&
		NOX-MT, Beacon, Maestro, Floodlight	&
		Throughput, Latency, Threading Scalability, Delay Sensitivity	&
		Switch Partitioning\newline Packet Batching\newline Task Batching	&
		Switch partitioning \& switch batching impacts throughput.\newline[3pt]
		Packet batching \& task batching impacts delay sensitivity.	\\\hline
		
		\cite{shalimov2013advanced}	&
		2 separate Xeon Servers\newline
		10 Gbps Link Speed	&
		CBench\newline Hcprobe	&
		NOX, POX, Floodlight, Ryu, Mul, \newline Beacon, Maestro	&
		Throughput, Latency, Reliability, Security	&
		Flow Modification\newline Customized Workload	&
		Scalability of controller depends on the number of cores.\newline[3pt]
		Not every controllers can handle heavy workload.	\\\hline
		
		\cite{jarschel2012flexible}	&
		Single testbed with 4 servers (dual core)\newline 
		100 Mbps Link Speed	&
		OFCBenchmark	&
		NOX, Floodlight, Maestro	&
		Round Trip Time\newline Send and Response Rate\newline Packet Processing Rate	&
		Implement Boost Libraries to handle Threads	&
		Transmitting larger flows helps in detecting congestion in networks.\\\hline
		
		\cite{jarschel2014ofcprobe}	&
		Not Specified &
		OFCProbe	&
		NOX and Floodlight &
		Impact of Fat-tree Topology\newline Load Balancing	&
		Java library is used to handle OpenFlow connections &
		Topology has an impact on flow processing time.\newline
		Efficient handling of switch depends on the characteristic of controller.\\\hline
		
		\cite{Khattak2014} &
		5 $\times$ Server with Core i5 CPU &
		CBench &
		Floodlight and OpenDaylight &
		Throughput, Latency, Failure	&
		Not Specified &
		Custom profile is proposed for CBench.\newline[3pt]
		Controllers may suffer from memory leakages.\\\hline
		
		\cite{khondoker2014feature} &
		Not Specified &
		Analytic Hierarchy Process (AHP)	&
		POX, Floodlight, OpenDaylight, Ryu and Trema &
		Virtual Switch Support, Modularity, Documentation, API Compatibility &
		Not Specified &
		Evaluation Method is Subjective\newline[3pt] 
		Testing process may effect the outcome.\\\hline	
		
		\cite{zhao2015performance} &
		Single Testbed with Quad-Core Xeon Server &
		CBench, Open vSwitch	&
		NOX, POX, Floodlight, Ryu, Beacon &
		Throughput, Latency, Threading Capability, Python Interpretation	&
		Python Interpreter, Hyper-Threading (HT)	&
		HT offers performance improvement for java-based controllers.\newline[3pt]
		Reliability, Trustworthiness, Usability, and Scalability should be considered equally.\\\hline
		
		\cite{isaia2016performance} &
		1 $\times$ Quad-core, 1 $\times$ One Octa-core Testbed &
		Mininet, Open vSwitch, Indigo vSwitch	&
		POX &
		CPU Utilization, Topology Impact, Ping Delay	&
		Not Specified &
		Number of switches impact the flow installation time\newline[3pt]
		Mininet utilizes maximum system memory.\newline[3pt]
		Initial Ping Delay is larger than average Ping Delay.\\\hline
		
		\cite{Mallon2016} &
		1 $\times$ Multi-Core, 1 $\times$ Many-Core Testbed\newline
		10 Gbps Link Speed	&
		CBench &
		NOX-MT, Floodlight, Beacon, Maestro &
		Latency, Throughput, Energy Consumption, I/O Threading Impact	&
		Floodlight Learning Switch\newline
		CBench Delay Parameter\newline Maestro Config File Modification	&
		Number of Switches and cores impact NOX's performance.\newline[3pt]
		CPU types and system architecture impact scalability.\\\hline
		
		\cite{salman2016sdn} &
		Single Testbed with Octa-core CPU\newline 10 Gbps Link Speed	&
		CBench	&
		NOX, POX, Floodlight, OpenDaylight, ONOS, Ryu, IRIS, Beacon, Maestro	&
		Latency, Throughput	&
		Not Specified	&
		Controller's SBI allows additional support for future Internet architecture \\\hline
		
		\cite{suh2017toward} &
		Dual Core Virtual Testbed &
		Open vSwitch, Cluster Testbed, HTTP Generator, REST Client	&
		OpenDaylight,  ONOS &
		Flow Installation Rate\newline Flow Reading Rate\newline Failover Time	&
		Controllers are customized for WAN environment &
		Size of a cluster has impact on flow installation rate.\newline[3pt]
		Failover Time of a controller depends on number of devices.\newline[3pt]
		Latency has significant impact on large-scale WAN.	\\\hline
		
		\cite{bholebawa2018performance} &
		Not Specified &
		Mininet, Open vSwitch, Traffic Generator	&
		POX and Floodlight	&
		Round Trip Delay, Average Throughput	&
		Not Specified	&
		Simple controllers better suited for configuration-related tasks.\newline[3pt]
		Feature-based controllers are good for performance-based tasks. \\\hline
		
		\cite{nguyen2018benchmarking} &
		2 Xeon Testbeds &
		OFCProbe &
		ONOS &
		Topology Discovery Time, Path Provision Time, ASYN. Msg. Process Time	&
		Not Specified &
		Number of links has equal impact as number of switches regarding performance.\newline[3pt]
		Reactive path provisioning time relies on length of corresponding path.\\\hline			
	\end{tabularx}
\end{table*}

\subsection{Why Benchmark a Controller?}
Prior to executing SDN-based operations, network administrators are required to verify whether available components can match their requirements to perform necessary tasks. Hence, evaluations related to data plane (vSwitchs, links, etc.) may include tasks such as measurement of flow table capacity, progressing times of OpenFlow messages, and bandwidth utilization, etc. 
Similarly, for the control plane it is equally essential to evaluate whether the controller is capable to efficiently manage the complete network, and utilize the capabilities of data plane to its maximum capacity. Although the fundamental function of a controller is flow management and installation, a number of different performance metrics can be used for its benchmarking. As there are numerous controllers available with different architectures and properties, it becomes extremely important to have a standard benchmarking criteria for evaluation.

In this regard, there are two basic requirements: a) a set of benchmarking metrics, and b) an efficient tool for bench marking test. In \cite{bhuvaneswaran2018benchmarking}, authors have presented a basic list of tests which should be conducted to evaluate the performance of a controller. However, there can be a number of other metrics which should also be used when benchmarking different controllers. Similarly, the tool used to perform the test in an emulated environment is critical.

\subsection{Existing Works \& Lessons Learned}
Prior to this article, \cite{tootoonchian2012controller,shah2013architectural,shalimov2013advanced,jarschel2012flexible,jarschel2014ofcprobe,Khattak2014,khondoker2014feature,zhao2015performance,isaia2016performance,Mallon2016,salman2016sdn,suh2017toward,bholebawa2018performance,nguyen2018benchmarking} use multiple techniques, tools, and testbeds to evaluate the performance of several SDN controllers including scalability, reliability efficiency, and robustness.

In Table \ref{tab:Related_Comparison}, we compile most of the existing works associated with the evaluation of the controller performance and the major findings. Majority of these works use CBench~\cite{cbench-web}, to evaluate the performance based on latency and throughput. In most cases, throughput mainly correlates with threading capability of a controller, regarding the number of flows it can process in a specified time slot. Some other works extend CBench to integrate support with the operating system's kernel and compilers like Java and Python. The aim is to improve threading scalability of a controller regarding system's I/O modules. Some works  include simulation-based environments where hosts and vSwitches are virtualized to evaluate the impact of topology on the performance of a controller. In these experiments, the load balancing functionality is extensively tested. Moreover, some works evaluate the reliability of the controller by generating vulnerable flows. Energy consumption has also been evaluated using  fat-tree or data-center topologies. Below we give brief description of some of the notable works.

Authors in \cite{tootoonchian2012controller} present CBench \cite{cbench-web} tool for evaluation of different controllers.
They perform multiple flow-based experiments using it to compare the effectiveness and performance of NOX-MT, a multi-threaded adoption of NOX controller with other controllers like NOX, Beacon, and Maestro. Despite showing a notable improvement in performance, NOX-MT fails to identify some of the limitations of NOX such as massive utilization of Dynamic memory allocation and redundant representation of multiple requests. 

In \cite{shah2013architectural}, authors compare four multi-threaded controllers (NOX-MT, Floodlight, Beacon, and Maestro) for architectural features like multi-core availability, controller impact on OF switch, packet batching, and task processing. Authors use CBench to compare these controllers based on their throughput and latency performance. In throughput mode, two scenarios are considered including a fixed amount of switches with an increasing number of threads and fixed threads with an increasing number of switches. Beacon shows better performance in these two scenarios due to its ability to use the multi-core and multi-threading functionalities. Besides, the dynamic changing of packet sizes allows Maestro to perform better in latency test.

Work in \cite{shalimov2013advanced} presented a framework named HCprobe to compare seven different SDN controllers: NOX, POX, Floodlight, Beacon, Ryu, MUL and Maestro. To compare the effectiveness of these controllers, the authors performed some additional measurements like scalability, reliability, and security along with latency and throughput. The testbed analysis presents some security vulnerabilities along with the reliability issues with MUL and Maestro controllers. On the other hand, Beacon, MUL, and Floodlight obtained minimum latency while Beacon performed relatively well in the throughput test. 

Analytic Hierarchy Process (AHP) is used in \cite{khondoker2014feature} to analyze POX, Floodlight, OpenDaylight, Ryu, and Trema based on multiple standards like virtual switch support, modularity, documentation, programming language compatibility and availability of user interface. According to calculation, Ryu was elected to be the most suitable controller based on requirements as mentioned earlier. However, the AHP method is subjective and changing of measurements or scenarios may lead to a different outcome. 

In \cite{zhao2015performance}, authors use multi-core and many-core testbeds to evaluate NOX, Maestro, Floodlight, Beacon on the aspect of multi-core utilization efficiency, performance scalability, and energy consumption regarding data center environments. The work emphasizes on existing controllers limitation in taking advantages of the concurrency in modern hardware. 

In \cite{salman2016sdn} the performance of well-known centralized and distributed SDN controllers has been studied using CBench. The results show that both MUL and Libfluid MSG (written in C) achieved the highest throughput under an increasing number of switches whereas python-based Ryu and POX obtained better score in latency mode. However, with the increasing number of threads, both Beacon and MUL performed better while python-based controllers failed to show satisfying performance. 

\begin{table*}
	\centering
	\caption{Classification of Benchmarking Metrics and Tool Capabilities}
	\label{tab:Benchmark_metrics}
	\tiny
	\setlength\tabcolsep{4pt}
	\begin{tabularx}
		{0.8\linewidth}{|>{\hsize=0.6\hsize}Z|>{\hsize=1.4\hsize}Y|>{\hsize=2.5\hsize}Y|>{\hsize=0.5\hsize}Z|>{\hsize=0.6\hsize}Z|>{\hsize=0.4\hsize}Z|}\hline
		
		\multicolumn{2}{|>{\centering\setlength{\hsize}{2\hsize}\addtolength{\hsize}{3\tabcolsep}}X|}{Measurable Metrics}	& \multirow{2}{*}{Description} & 
		\multicolumn{3}{>{\centering\setlength{\hsize}{1.5\hsize}\addtolength{\hsize}{2.4\tabcolsep}}X|}{Benchmarking Tools} \\\cline{1-2}\cline{4-6}
		Group	& Parameters	&	& CBench	& PktBlaster	&OFNet\\\hline
		
		\multirow{3}{*}{Throughput}	&
		Async Message Processing Rate	&
		\multirow{3}{=}{Determines number of flow requests a controllers can process per unit time. A processed request does not mean a successfully installed flow.}	&
		\ding{51}	&\ding{51}	&\ding{109}\\ \cline{2-2}\cline{4-6}
		& Sync Message Processing Rate	&	&\ding{51}	&\ding{51}	&\ding{109}\\ \cline{2-2}\cline{4-6}
		& Send and Response Rate		&	&\ding{53}	&\ding{109}	&\ding{51}\\ \hline
		
		\multirow{3}{*}{Latency}	&
		Async Message Processing Time	&
		\multirow{3}{=}{Denotes the delay or time duration between request from the vSwitch and response received back.}	&
		\ding{51}	&\ding{51}	&\ding{51}\\ \cline{2-2}\cline{4-6}
		& Sync Message Processing Time	&	&\ding{51}	&\ding{51}	& \ding{51}\\ \cline{2-2}\cline{4-6}
		& Round Trip Time	&	& \ding{53}	&\ding{109}	&\ding{51}\\ \hline
		
		\multirow{5}{*}{Flow Related}	&
		Path Provision Time (Proactive/Reactive)	&
		\multirow{5}{=}{Determines the efficiency of a controller to install flows, or measures which include communication between a source and destination.}	&
		\ding{53}	&\ding{51}	&\ding{51}\\ \cline{2-2}\cline{4-6}
		& Path Provision Rate (Proactive/Reactive)	&	&\ding{53}	& \ding{51}	&\ding{109}\\ \cline{2-2}\cline{4-6}
		& Flow Reading Rate	&	&\ding{53}	&\ding{53}	&\ding{109}\\ \cline{2-2}\cline{4-6}
		& Flow Installation Time	&	&\ding{51}	&\ding{51}	&\ding{51}\\ \cline{2-2}\cline{4-6}
		& Load Balancing	&	&\ding{53}	&\ding{109}	&\ding{53}\\ \hline
		
		\multirow{2}{*}{Topology}	&
		Topology Discovery Time/Size	&
		\multirow{2}{=}{Measures the capability to discover topology or change in topology. This also indirectly measure the SBI performance.}	 &
		\ding{53}	&\ding{51}	&\ding{51}\\ \cline{2-2}\cline{4-6}
		& Topology Change Time	&	&\ding{53}	&\ding{51}	&\ding{51}\\ \hline
		
		\multirow{4}{*}{Threading} & Thread Capability	&
		\multirow{4}{=}{Indicates the utilization efficiency of a controller regarding the OS and physical hardware resources.}	&
		\ding{51}	&\ding{51}	&\ding{51}\\ \cline{2-2}\cline{4-6}
		& I/O Impact	&	&\ding{51}	&\ding{51}	&\ding{53}\\ \cline{2-2}\cline{4-6}
		& Control Session Capability	&	&\ding{53}	&\ding{109}	&\ding{53}\\ \cline{2-2}\cline{4-6}
		& vSwitch CPU Utilization	&	&\ding{53}	&\ding{53}	&\ding{51}\\ \hline
		
		\multirow{5}{*}{Others}  & Forwarding Table Capacity	&
		\multirow{5}{=}{Miscellaneous parameters which can be measured for specific scenarios.}	&
		\ding{53}	&\ding{51}	&\ding{53}\\ \cline{2-2}\cline{4-6}
		& Ping Delay Time	&	&\ding{53}	&\ding{53}	&\ding{53}\\ \cline{2-2}\cline{4-6}
		& Energy Consumption	&	&\ding{53}	&\ding{53}	&\ding{53}\\ \cline{2-2}\cline{4-6}
		& Network Re-provisioning Time	&	&\ding{53}	&\ding{53}	&\ding{53}\\ \cline{2-2}\cline{4-6}
		& Controller Failover Time	&	&\ding{53}	&\ding{109}	&\ding{53}\\\hline
	\end{tabularx}
\end{table*}


\subsection{Benchmarking Metrics and their Impact}
In this section we present a detailed list of performance metrics that can be used to benchmark SDN controllers. Table~\ref{tab:Benchmark_metrics} outlines the grouping and description of each of these metrics. Some of these have also been identified by \cite{bhuvaneswaran2018benchmarking}, however, we have extended this list and grouped them to eliminate the confusion regarding terminology. Generic terms such as, throughput and latency can have significantly different meaning depending on measurement process.
Additionally, there can be other metrics to evaluate a controller, e.g. security, reliability, etc. However, we refer to them as non-measurable parameters which are more subjective in nature. We leave their classification as future work. The measurable parameters are grouped as following. 

\subsubsection{Throughput Metrics}
Throughput is usually measured as a rate for processing flow requests by the controller.
The important thing to note, is that it is not the flow installation time (path provisioning). From the test tools perspective, it is the number of packet\_in messages sent and the corresponding packet\_out mssages recieved per unit time. 
These requests could be synchronous or asynchronously coming from the vSwitches in real environment.

\subsubsection{Latency Metrics}
This group of metrics is measured in time units. Similar to throughput it only deals with the time between packets sent to controller and response received at the vSwitch. A number of factors  can effect the latency of a controller, including computation time require by the controller and link delay. 

\subsubsection{Flow Related Metrics}
These metrics deal with the complete path provisioning and flow installation. The primary difference between this and throughput is the complete path. Throughput only measures the rate from vSwitch to controller and back to vSwitch. However, complete flow installation requires installation of flow entries at other vSwitches along the path.
We group both rate and time variants of these parameters in the same category, along with load balancing capability of the controller.

\subsubsection{Topology Based Metrics}
The ability to detect or determine a topology including its type (single, linear, overlay and tree), size and number of integrated nodes altogether represent a vital aspect to evaluate the efficiency of a controller. Interaction with its southbound interface also plays a significant role in these metrics.

\subsubsection{Threading \& Session Metrics}
This set of metrics identifies controller competence with respect to utilizing the system architecture, hardware capabilities, and I/O units. Optimization of thread-based capabilities like multi-threading offers several advantages of task batching, event scheduling, process flows as groups and most importantly increases controller's flow processing time and rate.

\subsubsection{Miscellaneous Metrics}
Here we group other parameters which can also be used for evaluating the controllers. Some of these can be crucial in specialized scenarios. For example energy consumption in mobile environments where controllers are deployed on devices which are energy constrained. Similarly, in situations where hardware failure is a concern, the failover time needs to be reduced so that backup controllers can takeover as quickly as possible.

\section{Tools for Controller Benchmarking}
Evaluating or benchmarking the performance of a controller can be done either through simulation/emulation or by using a hardware based testbed. Although, hardware testbeds provide measurements which are closer to actual values in production environment, however their cost is significant for research community. Hence emulation based evaluations are common practice. However, for benchmarking of SDN controllers, the software tool used has to be extremely efficient and precise. In this section, we present a number of well known tools available for benchmarking, followed by analysis for their properties and benchmarking capabilities.


\subsection{Benchmarking Tools}
Following are some of the commonly used tools for benchmarking. Table~\ref{tab:Tool_comparison} provides a comparative analysis of the three main tools used for evaluation in this work.

\textbf{CBench}
\cite{cbench-web} is one of the fundamental benchmarking tools with open-source license. It is designed explicitly for evaluating the performance of OpenFlow SDN controllers which support OpenFlow 1.0 and 1.3. However, due to compatibility limitation, controllers with OpenFlow 1.3 may experience performance issues. There are two basic evaluation metrics in CBench, i.e., Latency and Throughput. To measure Latency, the vSwitch forwards a single packet\textunderscore in message towards the controller and waits for a response. Tests can be repeated several times to obtain the average performance. The total number of acknowledgments obtained in a test period is used to compute the average latency. As for throughput measurement, each vSwitch continuously sends as many packet\textunderscore in messages as possible, to estimate the capability of the controller.

\textbf{HCprobe}
\cite{shalimov2013advanced} is an open-source extension of CBench, developed with the combination of Python and Shell scripts, to provide additional performance evaluation capabilities, such as reliability and scalability. The emulated switch can send vulnerable OpenFlow messages to controllers to check for resiliency and trustability. Besides, the test engine utilizes a Linux kernel, which allows customizable and scalable tuning of CPU threading. This allows the tester to obtain more accurate performance statistics of an SDN controller. 

\textbf{WCBench}
\cite{wcbench_github} is another variants of CBench built in Python and utilizes the core library module of CBench. Compared to CBench, feature set of this tool goes beyond latency and throughput, and offers additional aspects of automated evaluation with detailed and graphical statistics. Although it extends the support of OpenFlow to version 1.3, the compatibility of WCBench is still limited to specific versions of ODL controller.

\textbf{OFCBenchmark}
\cite{jarschel2012flexible} is built using C++ and Boost library to address some of the limitations of CBench. The components of this benchmarking tool include a graphical dashboard (built with Delphi), virtualized scalable vSwitch which is the core module, and includes a client that can administer evaluation tests. The tool offers distributed benchmarking by allowing clients to run in multiple instances, and offers extensible benchmarking such as Round Trip Time (RTT), flow installation rate, and CPU utilization, etc.

\textbf{OFCProbe}
\cite{ofcprobe-web} is an upgraded version of OFCBenchmark which concentrates on maximizing the flexibility of SDN controllers by emulating a significant amount of OpenFlow switches in a large scale environment. It is re-designed using Java to make it a platform-independent tool and also to overcome the virtualization overhead caused by SDN emulation tool like Mininet \cite{mininet_web}. The core competence of this tool is to analyze the impact of the network topology during the evaluation executed by the client component.  

\textbf{PktBlaster}
\cite{pktblaster-report} is a unified test solution that emulates large scale SDN networks including network infrastructure and orchestration layers of SDN controllers. The free version with limited capabilities offers features such as, latency and throughput measurement with different testing profilies, i.e. TCP, UDP, ARP\textunderscore Request, and ARP\textunderscore Reply. A throughput test determines the rate at which the controller configures the flows in the switches. The latency test gives the exact time in milliseconds which the controller takes to process a flow in the switch. Although the free version is limited to 16 switches and 64 MAC address, it offers additional properties like Flow tables, Group tables, Meter tables, size of the Switch Buffer, and maximum entries per flow table.

\textbf{OFNet}
\cite{ofnet-web} is a combined approach to integrate OpenFlow network emulation with performance monitoring and visual debugging of SDN controllers. OFNet can be deployed in a system to generate different types of topologies. The inbuilt traffic generator produces different types of network traffic. It is capable to measure performance characteristics of the controller such as flow generations, flow failures, CPU utilization, flow table entries, average RTT, latency of flow setup, etc.

\begin{table}[!t]
	\centering
	\caption{Comparison of Benchmarking Tools.}
	\label{tab:Tool_comparison}
	\tiny
	\setlength\tabcolsep{2.5pt}
	\begin{tabularx}
		{\linewidth}{|>{\hsize=0.7\hsize}Z|>{\hsize=1.5\hsize}Y|>{\hsize=1.8\hsize}Y|>{\hsize=0.7\hsize}Z|>{\hsize=0.7\hsize}Z|>{\hsize=0.6\hsize}Z|}\hline
		
		\textbf{Tool} & \textbf{Advantages} & \textbf{Limitations} & \textbf{License} & \textbf{Availability} & \textbf{User Interface}\\\hline
		
		\textbf{CBench}	&
		Faster Analysis Execution\newline[3pt]
		Platform Independent\newline[3pt]
		Source Code is available	&
		vSwitchs limited to 256\newline[3pt]
		Supports only OpenFlow 1.0\newline[3pt]
		Flow Length is Limited\newline[3pt]
		Supports only IP-based traffic\newline[3pt]
		Lacks User-Interface	&
		Open-Source &
		Yes &
		CLI\\\hline	
		
		\textbf{PktBlaster}	&
		1000 Emulated Switches\newline[3pt]
		Customized Switch Groups\newline[3pt]
		Detailed Statistical Results\newline[3pt]
		Accuracy is better than CBench	&
		No Customized Topology\newline[3pt]
		No Application-based Traffic\newline[3pt]
		Free Edition lacks Deep Analysis	&
		Open-Source Proprietary	&
		Yes	&
		Web UI\\\hline
		
		\textbf{OFNet}	&
		In-depth Performance Analysis\newline[3pt]
		Self-defined Topology\newline[3pt]
		Various Traffic Profiles\newline[3pt]
		Flow Event Syntax\newline[3pt]
		Traffic Generator	&
		Benchmarks Relies on Topology\newline[3pt]
		Slower Test Duration	&
		Open-Source	&
		On Request	&
		GUI \\\hline
	\end{tabularx}
\end{table}


\subsection{Benchmarking Capabilities}
In this work we use the three of the tools, i.e. CBench, PktBlaster, and OFNet, to evaluate different controllers. It is important to note that none of the tools available can measure all performance statistics. In most of the previous works and the output of tools, the metrics are rather simplified. For example, the throughput of a controller can be interpreted in a number of different ways. Similarly, as shown in Table~\ref{tab:Benchmark_metrics}, the latency can be determined using different metrics. The columns on right side of table shows each individual metric which can be directly measured, indirectly measured, or not measurable by a specific tool. 

\section{Evaluation and Benchmarking of Controllers}
This section discusses performance of 9 different controllers using previously described benchmarking tools. To the best of our knowledge, no previous work has compared such a large number of controllers, and performed cross comparison using different tools. The controllers evaluated are NOX, POX, Floodlight, ODL, ONOS, Ryu, OpenMUL, Beacon, and Maestro. The reason to select these out of previously discussed 34, is the availability of controller source code or implementation. The 3 benchmarking tools used are CBench, PktBlaster, and OFNet. We use a virtualized environment to emulate the controller and tools in separate virtual machines, running on a 2.10 GHz i7-3612QM processor with 12 GB of DDR3 RAM. Ubuntu 16.04.03 LTS is the base operating system and 1 Gbps link connects the VMs.

It is important to note that all results are plotted as bar graphs. This is done to increase visual understanding of the reader. Overlapping nine different controller outputs in a single plot were not only visually confusing, but also made it difficult to infer any meaningful information.

\begin{table}[!t]
	\centering
	\caption{Parameters used in evaluation setup.}
	\label{tab:eval_parameters}
	\tiny
	\setlength\tabcolsep{3pt}
	\begin{tabularx}
		{0.7\linewidth}{|>{\hsize=0.5\hsize}Z|>{\hsize=1.5\hsize}Y|>{\hsize=1\hsize}Y|}\hline
		
			\textbf{Tool} & \multicolumn{2}{>{\centering\setlength{\hsize}{2.5\hsize}\addtolength{\hsize}{3.75\tabcolsep}}X|}{\textbf{ Parameter Values} } \\\hline
			
			\multirow{5}{*}{\textbf{CBench}} & Number of Switch & 2, 4, 8, 16 \\ \cline{2-3}
			& Number of Test Loops & 20 \\ \cline{2-3}
			& Test Duration & 300 sec \\ \cline{2-3}
			& MAC Addresses per Switch (Hosts) & 64 \\ \cline{2-3}
			& Delay between Test Intervals & 2 sec \\
			\hline
			
			\multirow{7}{*}{\textbf{PktBlaster}} & Number of Switch & 2, 4, 8, 16 \\ \cline{2-3}
			& Test Duration & 300 sec \\ \cline{2-3}
			& Number of Iterations & 5 \\ \cline{2-3}
			& Traffic Profile & TCP \\ \cline{2-3}
			& Ports per Switch (Hosts) & 64 \\ \cline{2-3}
			& Flow Counts per Table & 65536 (Default) \\ \cline{2-3}
			& Packet Length & 64 bytes \\
			\hline
			
			\multirow{5}{*}{\textbf{OFNet}} & Number of Hosts & 20 \\ \cline{2-3}
			& Number of Switchs & 7 \\ \cline{2-3}
			& Desired Traffic Rate & 100 flow/sec \\ \cline{2-3}
			& Flow measured by & Packet-out \& Flow-Mod \\ \cline{2-3}
			& Total Test Duration & 300 sec \\ 
			\hline
		\end{tabularx}
\end{table}

\subsection{Evaluation Setup}
Table\ref{tab:eval_parameters} shows the different parameters for evaluation setup. It is important to note that the programmable parameters in different tools are not identical, hence, we have tried to best possible extent to make them similar. However, once the parameters are set, all controllers use the same values.

\textbf{CBench} tests the performance by sending asynchronous messages. For latency the messages are in series, i.e. it send a packet\textunderscore in message to the emulated switch and waits for a response before sending the next one.
We execute 20 iterations with varying number of emulated switches to observe the impact of switches on the controller. On the other hand, with same parameters we test the throughput of the running controller. However, the packets are not sent in series, and requests are sent without waiting for a response. One execution, CBench outputs the flow messages a controller can handle per second. The results presented here are an average of number of responses per second from all switches in that execution.  

\textbf{PktBlaster} utilizes the in-built TCP-based traffic emulation profile that creates an OpenFlow session between the emulated switch and the controller. Due to free edition of tool the number of iterations is limited to 5. The nine controllers are evaluated based on latency (flow installation rate) and throughput (flow processing rate).   

\textbf{OFNet} uses a custom tree-based topology consisting of 7 emulated switches and 20 virtual hosts. We limit the number of hosts and switches due to limited resources available on emulating machines. Inbuilt traffic generator is used, which initiates and transfers multiple types of traffic, such as DNS, Web, Ping, NFS, Multi-cast, Large-send, FTP and Telnet among hosts in the emulated network much like Mininet Emulation environment. Host 2, 12 and 20 act as DNS, NFS and Multicast server respectively. We analyze metrics such as, Round Trip Time, average flow setup latency, vSwitch CPU utilization, number of flows missed by the controllers, number of flows  sent and received. OFNet provides analysis against time, hence the average of 10 iteration is plotted against a 300 seconds simulation.

\subsection{Latency Performance}
\begin{figure*}
	\centering
	\subfloat[CBench latency with varying number of switches.]{
		{\includegraphics[width=0.3\linewidth]{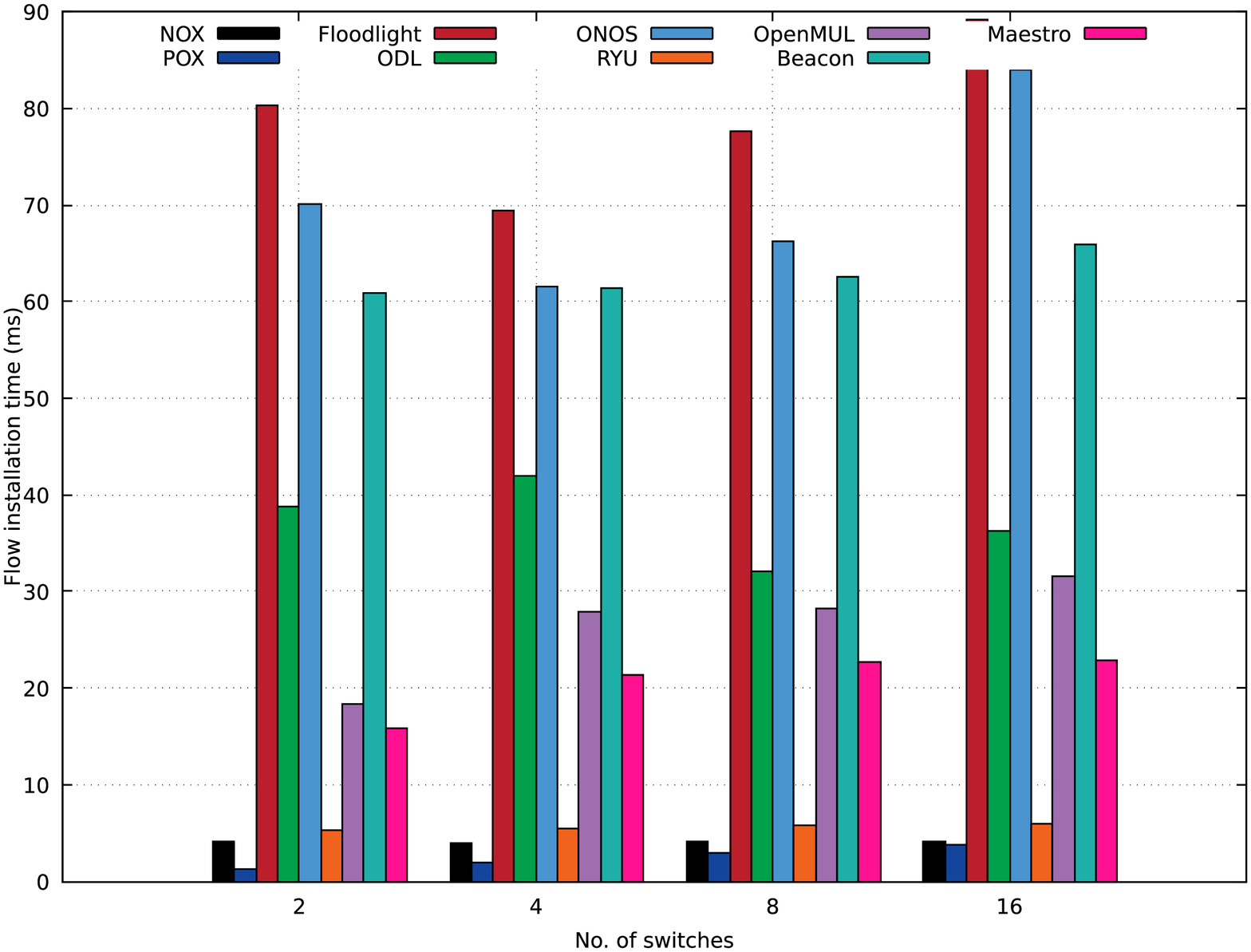}}
		\label{fig:latency_switch_cbench}
	}
	\hfil
	\subfloat[CBench latency in different number of iterations (16 switches).]{
		{\includegraphics[width=0.3\linewidth]{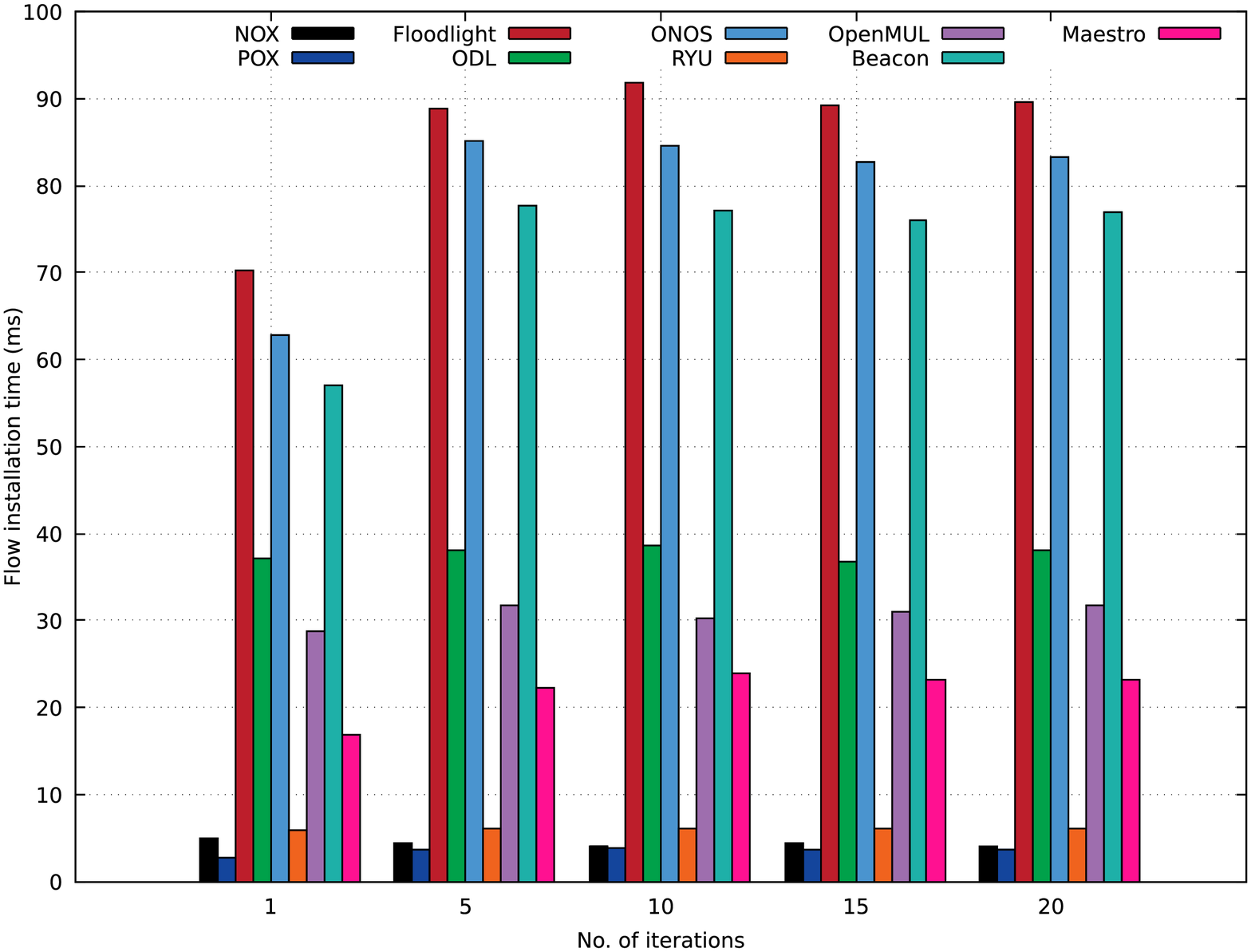}}
		\label{fig:latency_loop_cbench}
	}
	\\
	\subfloat[PktBlaster latency in with varying number of switches.]{
		{\includegraphics[width=0.3\linewidth]{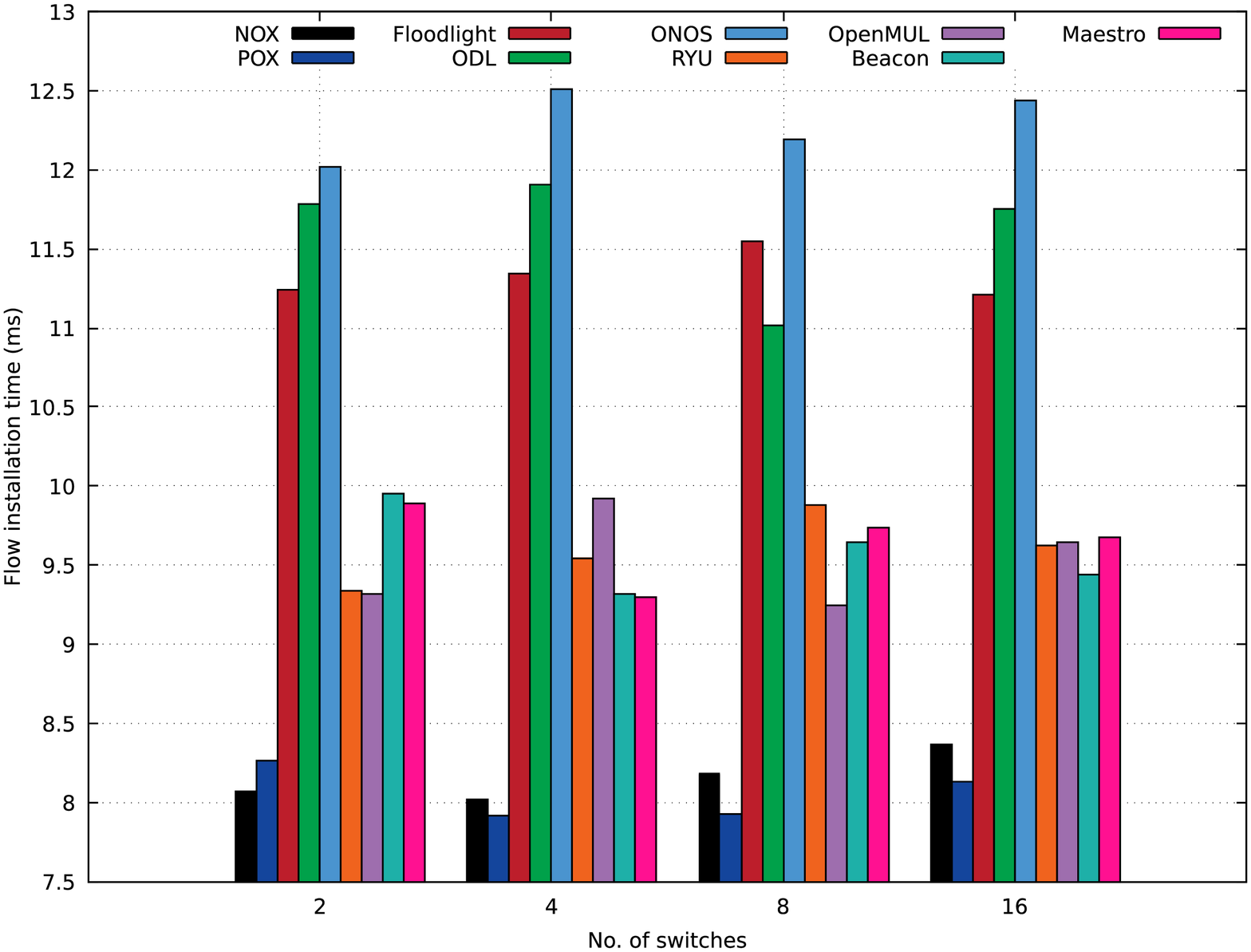}}
		\label{fig:latency_pktblaster}
	}
	\hfil
	\subfloat[OFNet flow setup Latency]{
		{\includegraphics[width=0.3\linewidth]{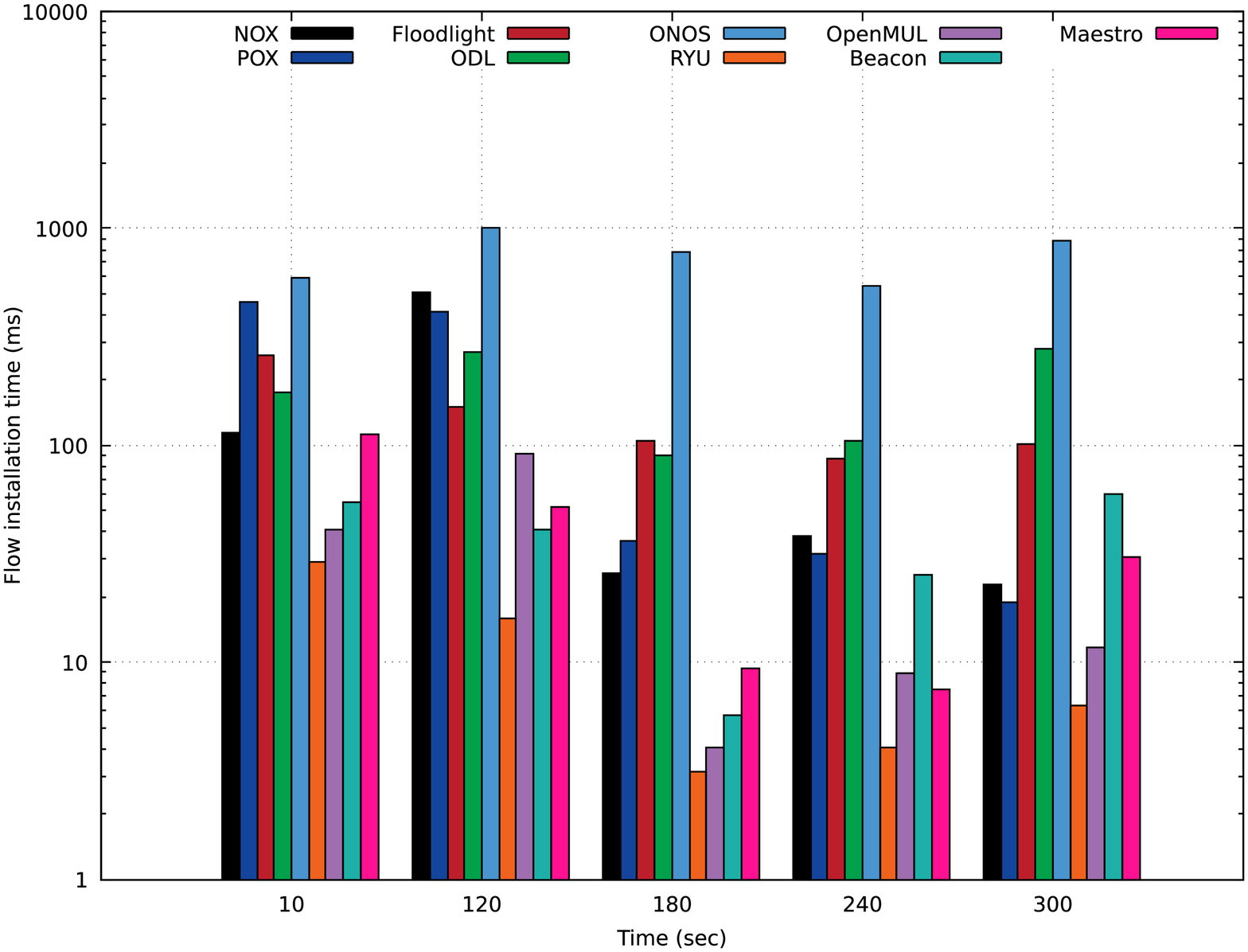}}
		\label{fig:flow_setup_latency}
	}
	\caption{Latency performance for CBench, PktBlaster, and OFNet.}
	\label{Fig:Latency_performance}
\end{figure*}
\subsubsection{CBench}
We observe two different effects on latency using CBench tool. First we observe the latency against number of switches in topology, from 2 to 16. Figure~\ref{fig:latency_switch_cbench} shows that there are two distinct groups, one with high latency, and one with significantly lower. An interesting observation is the Ryu controller which has negligible impact on its latency performance. Similarly, NOX and POX also show minimal change in latency as the switches increase. However, less latency does not translate to out-right winner, as the capabilities of controller itself must also be considered. In this regard, ODL, consistently performs in the middle and offers a number of other feature as listed in Table~\ref{table_controller_compare}.

The second experiment observes the effect of tool's own performance on latency measurement. Here we change the number of iterations while the number of switches is fixed at 16. Interestingly, the pattern in Figure~\ref{fig:latency_loop_cbench} shows most of the controllers to change their latency as the results are averaged-out over a larger set of repetitions. The basic take-away from this is that the setup environments effect on measurements should never be disregarded. It may positively or negatively impact the obtained results with the same parameters.

\subsubsection{PktBlaster}
Latency calculation using PktBlaster is also done against increasing number of switches. Figure~\ref{fig:latency_pktblaster} shows three distinct groups of controllers. NOX and POX show minimum latency, while Floodlight, ODL, and ONOS have the highest latency in this test. Ryu, OpenMUL, Maestro, and Beacon are in the middle. The important factor to note here is that the number of switches does not have any significant impact on the latency calculation. We again emphasis the fact, that the measurement process should reflect the metric being measured. Here latency is more closer to RTT between observing node and controller. On the other hand, flow installation time (path provisioning) would include multiple switches, hence increasing the time.

\subsubsection{OFNet}
Unlike CBench and PktBlaster, OFNet has a different evaluation and reporting method, where it simulates the SDN network much like Mininet. The output values are reported against time, instead of a specific value. Figure~\ref{fig:flow_setup_latency} shows the averaged result of 10 iterations on a time line of 300 seconds. It can be observed that there is no specific pattern over time followed by any given controller. The overall effect that we observe is that less time is required to install flows as the simulation progresses. The dip and rise in latency at around 180 sec mark is due to traffic generation artifact, where some types of traffic are generated later in the simulation, hence requiring more flows.

\subsubsection{Cross-Tool Analysis}
One of the contributions of this article is to demonstrate the difference in outcome for same metric under potentially similar network environments. As can be seen from Figure~\ref{Fig:Latency_performance} the Y-axis scale varies extensively in all three tools. For CBench the measured latency is in the orders of tens of milliseconds, where as in PktBlaster the same controllers perform under 10ms. In a total contrast the latency measurements on OFNet are in the order of hundreds of milliseconds. Controllers which performed the best in one simulator, are the worst performs in the other. Although OFNet has a different topological setup, however there is no correlation in the observed results.

\subsection{Throughput Performance}
This metric is measured using CBench and PktBlaster only as shown in Figure~\ref{Fig:Throughput_performance}. OFNet does not provide direct measurement of flow processing, however, indirect measurement can be done through sent and received flow messages, which is discussed in later section. 
\subsubsection{CBench}
In throughput mode, CBench switches send as many packets as possible at once, and does do not wait for a reply.  Figure \ref{fig:throughput_switch_cbench} shows the comparison based on increasing number of switches.  It is observed that NOX, POX, and RYU remain the lowest performers, while controllers like ODL, Beacon and Maestro have up to 100 responses per milliseconds. Although both OpenMUL and Floodlight performed consistently well around 150 flows/ms, the flow response rate of ONOS is significantly higher around 400 flows/ms to 500 flows/ms.

\subsubsection{PktBlaster}
\begin{figure*}
	\centering
	\subfloat[CBench throughput with varying number of switches.]{
		{\includegraphics[width=0.3\linewidth]{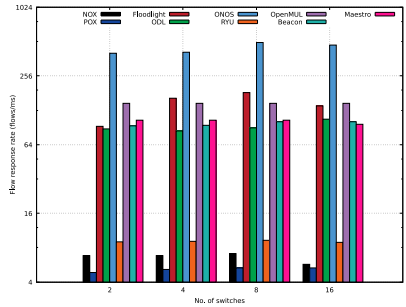}}
		\label{fig:throughput_switch_cbench}
	}
	\hfil
	\subfloat[PktBlaster throughput with varying number of switches.]{
		{\includegraphics[width=0.3\linewidth]{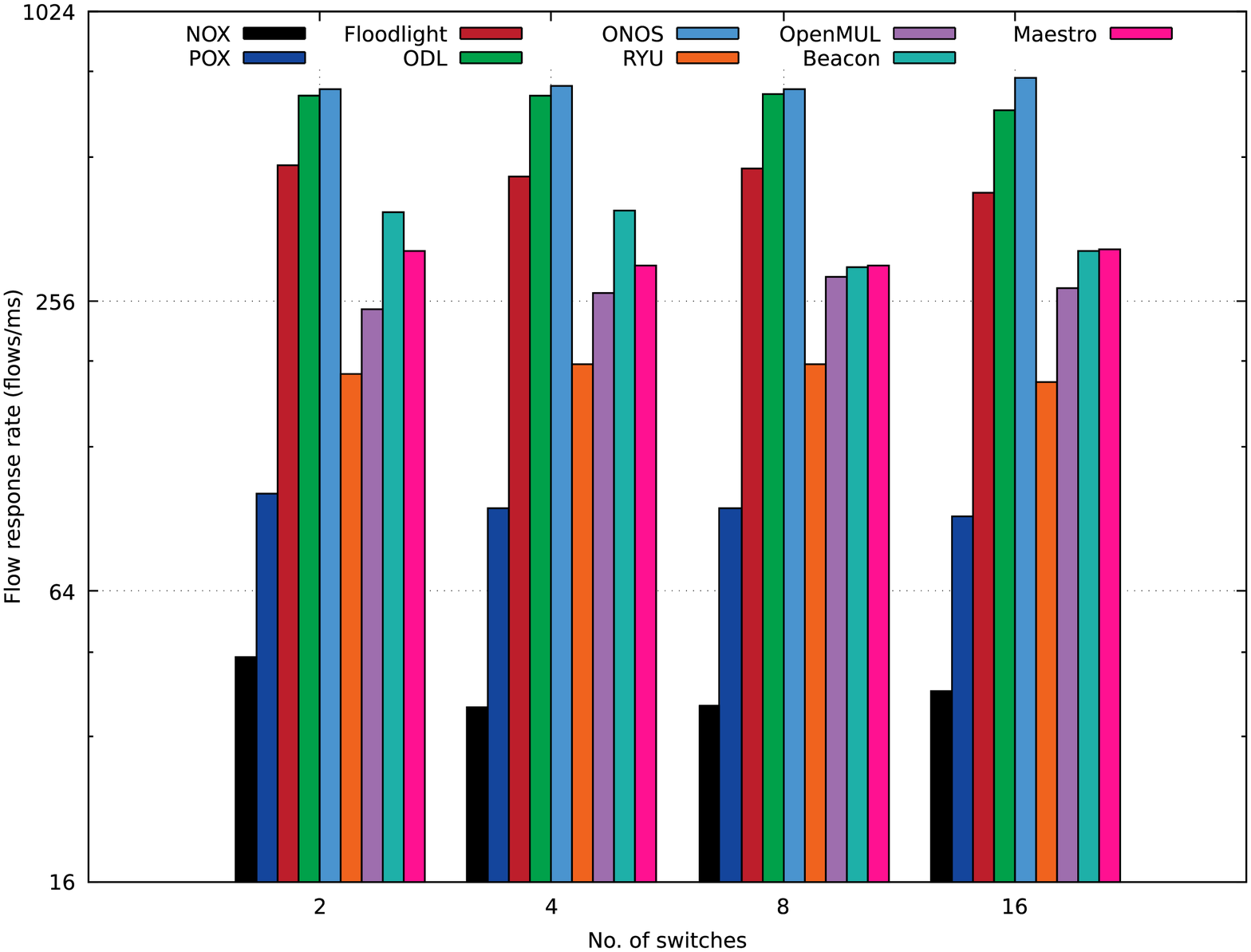}}
		\label{fig:throughput_pktblaster}
	}
	\caption{Throughput performance for CBench and PktBlaster.}
	\label{Fig:Throughput_performance}
\end{figure*}

The measurements of throughput shown in Figure~\ref{fig:throughput_pktblaster} present minimal effect from change in number of switches when testing with PktBlaster. The performance of Floodlight, ODL, and ONOS is the best among all the controllers compared, while NOX and POX are at the lower end. A minor (insignificant) decrease in throughput was observed as the number of switches increased for NOX, POX, and Ryu. However, after running 5 iterations each, the change remains insignificant.

\subsubsection{Cross-Tool Analysis}
Similar to earlier analysis, the tools differ in throughput metric also, however the change is not too drastic. All the controllers tend to perform better in PktBlaster evaluations as compared to CBench. Specifically, ODL and Floodlight show significant gain in the performance.

\subsection{OFNet Specific Measurements}
In this set of experiments, we focus specifically on the performance metrics offered by OFNet.
 
\subsubsection{Average Round Trip Time}
\begin{figure*}
	\centering
	\subfloat[Average RTT Measurement.]{
		{\includegraphics[width=0.3\linewidth]{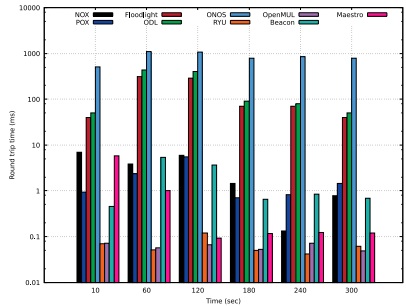}}
		\label{fig:rtt}
	}
	\hfil
	\subfloat[CPU utlization of vSwitch Daemon.]{
		{\includegraphics[width=0.3\linewidth]{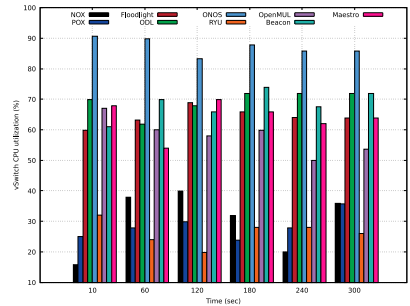}}
		\label{fig:switch_cpu}
	}
	\caption{RTT and CPU performance for OFNet.}
	\label{Fig:RTT_CPU_performance}
\end{figure*}

RTT evaluation is an important factor to consider when identifying the location of controller deployment. 
It identifies the communication delay between the controller and the switch. If the controller and switches are physically far apart, the increased RTT will contribute to increased latency. Similarly, the time complexity of packet processing at controller effects the overall performance. 
Based on our tree topology, Figure~\ref{fig:rtt} shows that ONOS has high RTT that starts with 100 ms and goes past 1000 ms during the simulation. On the other hand, Ryu \& OpenMUL have least RTTs, mostly because of less complex algorithms involved at the controller. However, less complex does not translate to better, rather, they may be attributed to less number of controller capabilities.

\subsubsection{CPU Utilization of vSwitch Daemon}
Here we use the OFNet's in-built traffic emulation application to transmit various packets and to identify the CPU usage of the vSwitch process while the vSwitch is interacting with a controller. While running a single-threaded controller like NOX, POX, and RYU, the CPU utilization in Figure~\ref{fig:switch_cpu} of vSwitch daemon remains under 30\% to 40\%. On the contrary, CPU utilization is remarkably higher at 90\% in the case of the multi-threaded controller like ONOS. Besides, the CPU usage remains under 70\%  rest of the controllers including Floodlight and ODL. One major factor in high throughput performance of ONOS is the multi-threading capabilities. However, they can be limited by the capabilities of the vSwitches.

\subsubsection{Missed Flows}
\begin{figure*}
	\centering
	\subfloat[Missing Flows.]{
		{\includegraphics[width=0.3\linewidth]{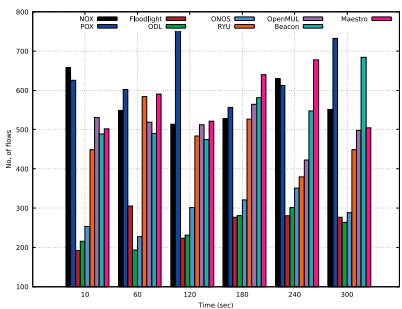}}
		\label{fig:missing_flow}
	}
	\hfil
	\subfloat[Flows Sent to Controller.]{
		{\includegraphics[width=0.3\linewidth]{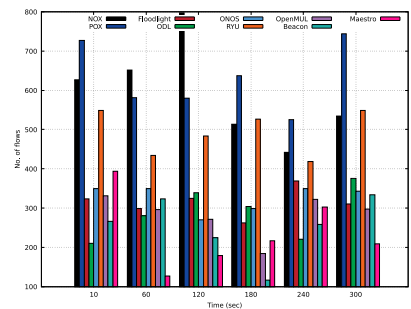}}
		\label{fig:of_sent}
	}
	\hfil
	\subfloat[Flows Received from Controller.]{
		{\includegraphics[width=0.3\linewidth]{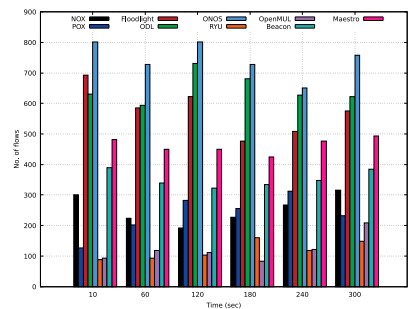}}
		\label{fig:of_received}
	}
	\caption{Flow measurements for OFNet.}
	\label{Fig:Flow_performance}
\end{figure*}

Here we measure the number of flows that the controller misses while the test is ongoing. Typically the traffic generator initiates flow requests to the vSwitches, which in-turn sends requests to the controllers and waits for the response. In this testing environment, vSwitch transmits reactive flows to benchmark the SDN controllers. Figure~\ref{fig:missing_flow} depicts that, ONOS, ODL and Floodlight miss the least number of flows as opposed to NOX, POX and RYU. This again is attributed to the multi-threading capabilities of the controllers, which allows them to perform comparatively better than the single-threaded ones.

\subsubsection{Flow Messages Sent \& Received}
This experiment calculates the number of flow messages that have been sent to the controller by vSwicth and the received flow messages from the controller. Although, both CBench and PktBlaster use the term "Packet\textunderscore in" to send flows towards controller to evaluate latency and throughput, OFNet instead sends flow messages continuously at a specific duration to determine the flow acceptance efficiency of the controller. Figure~\ref{fig:of_sent} shows that least amount of OF messages have been sent to NOX, POX, RYU, and OpenMul compared to others while a significant amount of messages has been transmitted to ONOS, ODL and Floodlight controllers. Figure~\ref{fig:of_received} depicts that, the flow reception rate is higher from the controllers like NOX, POX, and RYU as these controllers have less computational time. On the contrary, flow reception rate of multi-threaded controllers such as Floodlight, ONOS, and ODL is less than the single-threaded ones, which is due to the distributed nature of these controllers. As the received messages are coming from a specific instance of the controller, hence the plot reflects a lesser value.

\section{Research Findings}
Based on the qualitative analysis of controllers, properties \& capabilities of benchmarking tools, and the evaluation of controllers using them, we have summarized the main findings below.

\begin{itemize}
\item Considering latency and throughput, multi-threaded controllers including centralized ones (Floodlight, OpenMul, Beacon, Maestro) and distributed ones (OpenDaylight and ONOS) perform significantly better than centralized and single-threaded controllers like NOX, POX, and Ryu. However, they also require more physical resources in order to perform efficiently. 

\item Majority of the controllers proposed in literature have no implementation available and the details available are not sufficient for third person to code it. Hence, other than theoretical comparison, it is not possible to evaluate them.

\item Placement of controller in physical topology, directly impacts a number of performance parameters. In this regard, we plan to conduct an extensive study with different topological setups (datacenter, WAN, mobile, etc.) to compare distributed controllers.

\item Limitations of tools also directly effect the benchmarking. For CBench and PktBlaster we only utilized a specified number of the emulated switches due to available hardware resources and in-built traffic profiles. Therefore, physical resource and modification of compiler (or interpreter) may have some noticeable impact on the collected results.

\item We also noticed that some of the available features of tools, such as packet length, vSwitch buffer size, etc. impact the performance of the controller. However it is important to note that the outputs given by any tool also indicate the performance of components used in complete topology. Isolating the performance of controller from the results is not possible.

\item Utilization of benchmarking tool like OFNet allows us to define custom topology with a variety of traffic profiles. We observed that single-threaded centralized controller can still perform better in simplified topologies while multi-threaded controllers are more suitable for complex environments.
\end{itemize}
	
\section{Conclusion}
Benchmarking the performance of a controller is a challenging task. In this work we qualitatively compare 34 controllers, and then perform benchmarking and evaluation in quantitative terms for 9 controllers. During this process, we have also categorized and classified the different metrics which should be used for controller benchmarking. Moreover, we conduct an analysis of tools which can be used in the benchmarking process. Based on the observations, we find that very few controllers comply to OpenFlow 1.3 (or higher version) and provide enough information for actual deployment. Most of the evaluations done previously are based on simple metrics , with specific optimization objectives. Moreover, the tools used vary significantly in features and capabilities. It is impractical to compare results of one tool with another. Simulation/emulation based evaluation can give only an indication of performance at best, and may significantly differ from actual production environment evaluation.

\bibliographystyle{IEEEtran}
\bibliography{references}
\end{document}